\documentclass[a4paper]{article}
\usepackage[margin=2cm]{geometry}
\usepackage{eurosym}

\title{Challenges for Musical Education\\
	in the Age of AI and Digital Transformation}
\author{Jean-Pierre Briot \\[5pt]
{\em Sorbonne Universit\'e, CNRS, LIP6\footnote{{\tt https://perso.lip6.fr/Jean-Pierre.Briot/}.},
	F-75005 Paris, France}\\
{\em DI/PUC-Rio\footnote{{\tt https://www.inf.puc-rio.br/blog/professor/jean-pierre-briot/}.},
	Rio de Janeiro, RJ 22451-900, Brazil}\\
{\em CBAE/UFRJ\footnote{{\tt https://cbae.ufrj.br/programa-de-catedras/programa-de-catedras-2023/\#:\~:text=Criatividade}.},
	Rio de Janeiro, RJ 22250-020, Brazil}}
\begin{document}

\maketitle

\abstract{Recent and profound technological changes
(digital audio workstations, streaming, and generative AI)
compel us to rethink how music is created, listened to, and -- consequently -- taught.
In this article, we analyze the different types of change and their effects on the transformation of the music ecology
and economy, as well as on the lives of musicians.
We also examine the consequences for education and outline possible future directions.}

\begin{center}
\end{center}

\section{Introduction}

Music education has never been a static discipline. Each major technological shift\footnote{From
	the printing press that democratized musical scores, to the phonograph that first separated listening from live performance, to electric amplification that reshaped the sonic landscape.}
has forced educators and institutions to reconsider what they teach, how they teach it, and why. We now stand at what may be the most consequential of such turning points.
Three deeply intertwined transformations have been converging simultaneously:

\begin{itemize}
\item The very {\em nature} of music has changed: how it is made, distributed, consumed, and valued;
\item The {\em public} for music has changed: listening habits are now shaped by streaming algorithms and the boundary between consumer and creator has blurred;
\item Music-{\em making} itself has changed: digital audio workstations (DAWs) have for two decades been reshaping compositional practice.
In addition, generative AI has now irrupted, capable of producing complete, stylistically coherent musical pieces from a short text prompt.
\end{itemize}

These changes are not independent of one another, and they all bear directly on musical education -- both the content that must be taught, and the pedagogical tools and methods available to teach it.
This paper attempts to map these challenges and to consider how education might adapt. Section~\ref{sec:music-changed} surveys the changes in various aspects of music (nature, production, public, economics).
Section~\ref{sec:education-change} examines the implications for education before Section~\ref{sec:conclusion} concludes.

\section{Music Has Changed}
\label{sec:music-changed}

\subsection{The Nature of Music Has Changed}

\subsubsection{From Instrument to Stream: A Chain of Technological Revolutions}

The 20th century has produced a succession of technological revolutions that have progressively altered the nature of the musical artifact. The electric microphone and the recording industry have decoupled musical performance from the concert hall.
The phonograph, radio, and then television gave music unprecedented reach, although the artifact still remained physical: a vinyl disc, a magnetic tape, a compact disc.
Each of these technologies carried its own economic model, its own set of skills to acquire, and its own set of gatekeepers.

The MP3 format (standardized in 1991, popularized in the late 1990s) freed music from its physical carrier, opening the era of digital dematerialization.
The consequences were immediate and severe for traditional music industry: revenues from recorded music collapsed in the 2000s.
Streaming platforms (like Deezer, launched in 2007, and Spotify in 2008) proposed a new model: unlimited access in exchange for a subscription or attention (advertising). By 2024, streaming accounts for the overwhelming majority of recorded music revenues in most Western markets.

\subsubsection{Dematerialization and Its Consequences}
\label{sec:dematerialization}

Dematerialization does not mean that music has become intangible in a purely technical sense -- digital files are perfectly reproducible material objects.
What has dematerialized is \emph{scarcity}. As Fran\c{c}ois Pachet has argued, the transition from physical to digital artefact implies a transition from uniqueness and rarity to ubiquity and value dilution \cite{Pachet2025}.
The same observation applies to music production tools: a fully equipped recording studio, which once required capital investments of hundreds of thousands of dollars, can today be approximated by a laptop, a microphone, and open-source software.

This creates what Pachet calls the \emph{paradox of creators}\footnote{``Artists today
	face a profound and increasingly complex paradox: they want their creations to be heard, shared, and experienced by as many people as possible,
	but they also want to keep their work uniquely theirs, resisting the idea that it could be copied or used to train algorithms that mimic their style.'' \cite{Pachet2025}.}:
digital technologies and the Internet have simultaneously given creators unprecedented reach and diffusion -- a musician in a small town can now find a global audience -- while structurally destroying the economic value of the very content they help distribute.
Better diffusion and lower value are two sides of a same coin. The streaming economy exemplifies this paradox acutely: a track can accumulate millions of plays and its creator still earns less than the price of a coffee.
Generative AI is now further deepening this paradox, flooding distribution channels with machine-produced content at near-zero marginal cost and further compressing the attention and revenue available to human creators.

This has a direct educational implication: technical mastery of physical instruments and the social prestige it commands is no longer a sufficient differentiator.
The barriers to creating and distributing music have fallen, but the barriers to being \emph{heard} and \emph{valued} have in some ways risen sharply.
Music education must prepare students for this uncomfortable duality -- equipping them with the tools to reach audiences widely, while helping them develop the artistic distinctiveness and economic literacy needed to sustain a creative life in an environment of hyperabundance.

\subsubsection{Democratization of Creation}

Paradoxically, dematerialization has enabled an unprecedented \emph{democratization} of music creation. Platforms like SoundCloud, Bandcamp, YouTube, and TikTok allow anyone to distribute music globally with near-zero marginal cost.
The number of tracks added to Spotify in average per day exceeded 100,000 by 2024, with estimates suggesting that roughly half of this volume has now an AI-assisted or fully AI-generated component.

This democratization is genuinely liberating in some respects: styles and traditions previously confined to specific regions or communities can now find global audiences.
Bedroom producers from Rio de Janeiro, Lagos, or Seoul can reach listeners worldwide. The cultural diversity of music available to any listener has, in principle, never been greater.

But democratization of production also means \emph{saturation}.
The relationship between creator and audience -- the social contract that once made music economically and culturally meaningful for both parties -- is under strain.
Music education must now grapple with training musicians not merely for a craft but for a landscape where {\em attention} is the scarcest resource.

\subsubsection{Impact on Rights and Revenue}

The streaming economy has created winners and losers whose distribution is highly unequal.
A very small number of artists -- and above all the major record companies that hold their rights -- capture a disproportionate share of streaming revenues, while the vast middle of working musicians earns very little per stream.
The per-stream royalty on major platforms is fractions of a cent.

Generative AI has still further complicated the rights and revenue landscape. Systems such as Suno and Udio have been trained on vast corpora of existing recordings,
raising unresolved questions about copyright, consent, and fair compensation for the artists whose works constituted those training sets.
In October 2025, a significant agreement has been reached between Universal Music Group and Udio, a first step toward new legal frameworks.
But the jurisprudence remains in rapid evolution, and music educators must prepare students to navigate in an environment where the rules are still under evolution.

\subsection{The Music Public Has Changed: The Rise of Functional Music}

One of the subtler but more significant shifts of the streaming era is a change in \emph{how} people listen.
Streaming platforms have actively shaped listening habits through algorithmic recommendation, nudging users toward music suited to functional contexts: studying, exercising, sleeping, relaxing, focusing.
The result has been a significant growth in what might be called \emph{functional music} -- music valued primarily for its effect (concentration, relaxation, mood, etc.)
rather than for the aesthetic and expressive qualities that have historically been the focus of musical education\footnote{That said,
	we should remember that music had historically developed for various social purposes (for example, religious rites or dance).}.

This shift has both quantitative and qualitative dimensions. Quantitatively, studies of popular Western music over the past five decades have documented a restriction in melodic variety, a homogenization of timbre, and an increase in dynamic compression \cite{Serra2012},
consistent with the optimization of music for background listening and for streaming platforms' sonic environments.
Qualitatively, a tension is emerging between music conceived as autonomous artistic expression and music conceived as functional content -- a distinction with profound implications for what we choose to teach and what we choose to value.

\subsection{Music Making Has Changed}

\subsubsection{Digital Audio Workstations (DAWs)}

The widespread adoption of DAW software -- Ableton Live, Cubase, GarageBand, Logic Pro, and others -- over the past two decades has already transformed music production practice.
The DAW workflow is fundamentally different from traditional composition.
Where traditional composition is {\em top-down} and {\em score}-centered -- beginning with a melodic or harmonic inspiration, refined through notation, then orchestrated --
DAW-based production is {\em bottom-up} and {\em sound}-centered\footnote{Although
	symbolic melody, chords, voicing or rhythmic tracks, imported or exported in the MIDI format, are also often used.},
beginning with loops, samples, and sonic textures that are assembled, layered, and manipulated.

This distinction is not new: Pierre Schaeffer's ``musique concr\`ete'' in the 1940s already operated by assembling and transforming recorded sounds\footnote{Using
	vinyls records with ``sillons ferm\'es'' and then looped tapes.}.
The digital affordance of non-destructive editing, instantaneous duplication, and near-infinite layering has made this approach the dominant mode of popular music production.
Students who come to music education today may have spent years in a DAW environment and may have developed a sophisticated intuitive understanding of timbre, texture, and production aesthetics
-- skills that traditional conservatory curricula were not designed to recognize or develop.

\subsubsection{AI in the Picture}
\label{sec:ai-in-picture}

The arrival of generative AI adds a further, qualitatively different layer to this transformation.
One must distinguish, however, between two very different approaches to AI in music, which correspond to different purposes and different relationships between the \emph{human creator} and the \emph{machine}.

\begin{itemize}
\item \emph{Autonomous generators} -- Systems such as Suno and Udio -- produce complete, stylistically coherent musical pieces from a text prompt.
The audio quality of these systems has progressed remarkably fast. In terms of sonic production, results that would have been unimaginable even five years ago are now generated in tens of seconds.
These systems are well suited for \emph{functional music} -- background music for videos, advertisements, and documentaries -- where speed, cost, and stylistic competence matter more than compositional originality or artistic voice.
\end{itemize}

However, these systems have clear and fundamental limitations from a musical and educational standpoint.
Their generation is \emph{end-to-end}: a complete regeneration is required for any structural modification\footnote{At the time
	of writing, only remixing through regeneration is possible.},
and there is no lever for intervention at the \emph{compositional level} -- melody, harmony, and structure are opaque and uncontrollable.
More fundamentally, as Briot \cite{Briot2025} argues, these systems learn from existing musical artefacts; they do not model the \emph{process} of creation.
They are statistical interpolators over a learned stylistic space\footnote{See, e.g.,
	\cite{Briot2021} for some historical introduction to music generation with current AI techniques.},
which structurally biases them toward conformance and away from originality.
Some further risk is what Shumailov et al. \cite{Shumailov2024} call \emph{model collapse}: as AI-generated content floods training datasets, successive generations of models converge to lower and lower variance through this retro-alimentation.

\begin{itemize}
\item \emph{Composition assistants} represent a fundamentally different philosophy: rather than \emph{replacing} the composer(s), they support \emph{interactively} their practice and operational steps.
Two landmark systems in this lineage, both developed by Fran\c{c}ois Pachet and his collaborators at Sony CSL-Paris, illustrate the range of what this interaction can mean.
\end{itemize}

FlowComposer, developed within the ERC Flow Machines project \cite{Pachet2021}, is a score-centered composition assistant.
The musician can interactively write and edit a musical score, and at any point solicit FlowComposer for melodic or harmonic sketches inspired by a chosen stylistic corpus -- e.g., the bossa novas of Tom Jobim or the jazz standards corpus --
as well as completions and harmonizations of melodic fragments they have already written.
Crucially, constraints can be imposed on the generation: e.g., ending a melody on the same note as the beginning, avoiding excessively long notes, conforming to a specified harmonic structure\footnote{Note that
	FlowComposer as well as Continuator are based on Markov models and not current deep learning (artificial neuron networks) models, which are lighter, although less conformant to the style learnt,
	and easier to integrate with constraints (see, e.g., a comparison in Section 1.2.3 of \cite{Briot2020}).}.
The composer retains full intentionality and control over the compositional refinement process as well as the structural and aesthetic decisions.
FlowComposer has been validated in the creation of the album Hello World in 2018, a landmark demonstration of AI-assisted composition in which the system was subordinated throughout to the artistic vision of various professional musicians.
Final production as well as actual instrument playing and singing were almost only performed by human musicians, starting from the composition and pre-production demo constructed with the help of the FlowComposer environment.

The Continuator \cite{Pachet2002} addresses a different but complementary dimension of musical creativity: \emph{real-time interaction} and \emph{improvisation}.
Where FlowComposer assists in the iterative process of composing a score, the Continuator operates at \emph{performance time}.
The system learns the style of a player from what they have just played, and immediately continues in that style -- responding to the musician as a stylistically attuned partner would in a jazz jam session, picking up where the human left off and extending their musical thought.
This makes the Continuator a tool for exploring improvisation, which might be understood as real-time instant composition: a mode of musical creativity that is spontaneous, embodied, and fundamentally interactive.
Experiments with professional musicians -- but also with children \cite{Rowe2018} -- showed that the system generated strong engagement, surprise, and a heightened sense of musical agency --
reactions that are educationally significant precisely because they arise from the experience of being musically \emph{heard} and \emph{answered} in real time.

The distinction between autonomous generators and composition assistants matters enormously for musical education.
The former positions students as consumers of AI output (or at best curators of it); the latter positions AI as an extension of the student's own creative agency. Only the second approach is educationally productive in the full sense.

It is also worth noting that \emph{audio} versus \emph{score} is still some fundamental divide in how AI systems operate on music.
Systems like Suno and Udio operate on raw audio -- a waveform, rich in timbre and texture but opaque to compositional analysis.
Systems like FlowComposer operate on symbolic representations -- notes, chords, durations -- which are interpretable, controllable, and educationally meaningful.
Some yet distant Graal for next generation AI-based music systems would be to manage simultaneously symbolic and audio contents and therefore offer the compositional transparency of symbolic systems combined with the stylistic richness currently found only in audio systems.

\subsection{Economics of Music and Musicians has Changed}

\subsubsection{The Streaming Major Disruption}

It is important to be precise about chronology and causality, because the temptation to attribute the financial precarity of musicians to generative AI is understandable but historically inaccurate.
The deep transformation of musicians' economics was already well underway, and largely complete, before generative AI became commercially relevant.
As discussed in Section~\ref{sec:dematerialization}, the shift from physical sales to streaming subscriptions, arriving at the end of the 2000s, restructured the economics of recorded music from a model based on unit sales (a fan pays for a copy)
to a model based on aggregated, pooled, and redistributed subscription revenue (a fan pays for access to everything, and the pool is split according to relative share of total listening).
This shift, on its own, was sufficient to collapse the per-unit value of a piece of recorded music by roughly two orders of magnitude relative to the sale of a CD or a download.

The numbers from the streaming era alone are stark, independent of any AI consideration.
Per-stream payouts remain a small fraction of a cent on most major platforms -- on the order of \$0.003 to \$0.005 on Spotify, somewhat higher on Apple Music and Tidal, and lower still on YouTube Music's ad-supported tier.
An artist needs on the order of hundreds of thousands of streams to generate revenue equivalent to a single modest gig.
While the aggregate sums paid out by platforms have grown substantially -- Spotify alone reported some \$11 billion in royalty payouts in 2025 \cite{Spotify2026}
-- this growth has been highly concentrated: a small number of artists at the top of the distribution capture an outsized share, while the long tail of working musicians sees per-stream income that is, for practical purposes, negligible\footnote{That said,
	thanks to its growing revenue, recent Spotify numbers show more promising numbers \cite{Spotify2026}.
	In 2015, the 100,000th highest-earning artist generated in annual royalties from Spotify alone about \$350 and in 2025 more than \$7,300.
	Also, in 2015, only the very top artist on Spotify reached \$10 million in annual royalties from Spotify alone for the first time.
	Today, there are 80+ top artists who have reached this level and more than 1,500 artists have generated over \$1 million.
	The report states also that capturing just 1\% of streams from 1\% of listeners -- a small fraction of a small fraction -- is enough to earn \$1 million in annual royalties from Spotify.}.
Streaming has become, for most musicians, a discovery layer rather than an income source -- a ``welcome sign at the front door'' rather than the substance of a musical career.
This restructuring of musical value -- and the broader dynamic, discussed in Section~\ref{sec:dematerialization}, by which dematerialization converts what was once scarce (a physical recording) into something abundant and nearly free -- predates generative AI by well over a decade.

Generative AI did not \emph{initiate} this dynamic; it prolongs and intensifies it.
By making the production of stylistically competent recorded music nearly free at the margin, AI adds a further wave of supply into an already saturated and structurally devalued market for recorded music
-- over 100,000 new tracks are added to Spotify each day, with a substantial and growing share generated wholly or partly by AI.
The structural problem was already that recorded music had become abundant and cheap to access; AI's contribution is to make it abundant and cheap to \emph{produce} as well, compounding rather than originating the devaluation.
Understanding this sequence matters for music education, because it locates the more fundamental challenge correctly: even in a hypothetical world without generative AI, the economics of being a working recording musician would already be exceptionally difficult.
AI sharpens an existing crisis; it did not create it.

\subsubsection{Rights and the Economics of Music in the AI Era}

Within this already-transformed economic landscape, generative AI introduces a further complication that is qualitatively new: a copyright regime poorly suited to the kind of competition AI poses.
The core problem is structural: generative AI systems such as Suno and Udio typically do not reproduce a specific, identifiable sequence from a specific prior work.
They generate new audio that is statistically consistent with the styles, textures, and structures present in their training data, without (in most cases) directly copying an identifiable melody or sample from an identifiable source.
This is precisely what makes the existing copyright apparatus ill-suited to the problem: copyright law as currently constituted is designed to protect specific, fixed expressions -- a particular melody, a particular recording -- against direct copying.
It was never designed to address systems that absorb the statistical signature of an entire genre, an entire artist's catalogue,
or an entire era of music, and reproduce that signature in endless new variations, none of which infringes any single work, while collectively eroding the market for all of them.

This is, in effect, a new kind of harm that does not fit cleanly into the infringement categories the law already has tools for.
The training-data question (was it lawful to copy millions of songs onto a server to train the model?) is being litigated and, as of 2026,
gradually resolving toward negotiated licensing deals between major labels and AI companies -- Universal Music Group's and Warner Music Group's 2025 settlements with Udio and Suno being notable examples.
But this resolves only the upstream question of whether training was lawful; it does little for the downstream musician whose stylistic signature, technique, or niche can now be approximated by a generative system without any single act of copying that a court could point to.
Meanwhile, current copyright doctrine in most jurisdictions denies copyright protection to purely AI-generated output, on the grounds that it lacks sufficient human authorship\footnote{Note that
	the boundary between AI-generated and AI-assisted is not well defined, and different jurisdictions are reaching different conclusions.}
-- which means a musician who finds their style emulated by an AI system has, in most cases, no specific infringement to point to, while the resulting AI-generated competing work is itself currently unprotectable by anyone.
The legal categories simply were not built for a technology that competes by \emph{extrapolation} rather than by \emph{copying}.

We should also mention some positive prospects in the use of advanced technologies and AI (in addition to expand possibilities for artists, as exemplified in Section~\ref{sec:ai-in-picture})
to help composers and musicians in the registration and certification of their musical creations (composition and/or recording rights) as well at facilitating licensing and royalties management processes\footnote{Examples
	of such support platforms are URights by SACEM (French Society of Authors, Composers and Publishers of Music) in France and CertCon by Cedro Rosa Digital in Brazil.}.
Examples of the use of AI are in the automatic detection of plagiarism and then automatically processing for due royalties, and automatic detection of (only by) AI-generated music in streaming platforms (some facility currently proposed by Deezer).

For music education, these questions are not merely academic. Students need to understand the legal environment in which they will work. They need to know that using AI tools to generate music may complicate their ownership of the resulting work.
They need to understand how to document their creative process in ways that establish their human contribution. And they need to engage critically with the broader question of what fair remuneration for musicians should look like,
in a market reshaped first by streaming and now further reshaped by generative AI.

\subsubsection{What Remains for Musicians: Diversification and the Question of Collective Response}

Faced with both the structural decline of recorded-music revenue and a regulatory environment still catching up to generative AI, musicians and analysts of the music economy converge on a similar answer:
no single revenue stream is sufficient, and a sustainable musical livelihood in 2026 is necessarily a portfolio of several income sources, of which streaming is typically the smallest and least reliable.
Four of these deserve particular attention, because they share a common feature: they all depend on qualities that AI cannot replicate -- physical presence, direct relationship, and pedagogical authority.

\begin{itemize}
\item \emph{Other media for licensing authorial music} are used by some of the music composers: TV channels (for series, documentaries, etc.), movies, commercials.
Revenue is usually better than through streaming, but as for streaming, only a minority of composers can receive sufficient revenue.
Moreover, TV networks and producers are now often imposing so-called ``buy-out'' contracts (or strict flat-fee arrangements), consisting in one-off sum upon commissioning the work,
and in return the composer is required to relinquish a portion of their publishing rights.

\item \emph{Live performance} remains, by most accounts, the most resilient revenue source for working musicians.
A concert is an embodied, place-bound, time-bound event;
an AI cannot physically stand on a stage, and recorded or streamed substitutes have so far failed to replicate its economic and experiential value.
For many musicians, recorded music functions today primarily as a promotional vehicle for touring revenue, rather than the reverse.

\item \emph{Direct relationships with a niche of supporters}, built through platforms such as Patreon, Bandcamp, Ko-fi, or similar fan-subscription and direct-sales tools, allow musicians to convert a relatively small but genuinely engaged audience into stable income
-- exclusive content, early access, behind-the-scenes material, personal interaction.
This model explicitly trades the pursuit of mass algorithmic reach (where AI-generated content competes directly and at near-zero cost) for the cultivation of a smaller community bound by authentic relationship and identification with a specific human artist
-- a quality that, by construction, an anonymous or AI-generated alternative cannot offer.

\item \emph{Teaching and music education itself} -- private lessons, workshops, masterclasses, online courses --
recurs constantly in analyses of musician income as one of the most stable and accessible revenue streams, particularly for musicians who do not wish to tour constantly or who are between other engagements.
This observation closes a loop that runs through this entire paper: at the very moment that AI is disrupting the economics of music creation and distribution,
it is also expanding the toolkit available to musicians who teach, and the demand for musical mentorship may, somewhat paradoxically, be reinforced precisely because human teaching offers something increasingly scarce in an AI-saturated content economy
-- direct, accountable, {\em personally invested} transmission of musical knowledge from one person to another.
Music education is not only a domain that AI is transforming; for a great many musicians, it may also become an economic refuge from the very disruption that AI is causing elsewhere in their professional lives.
\end{itemize}

None of this amounts to a solution to the structural problem of declining recorded-music value
-- diversification is a survival strategy for individuals, not a fix for the economics of the industry as a whole, and it works far better for musicians with the time, digital literacy, and existing audience to build several income streams simultaneously than for those without these advantages.
This raises a question beyond the scope of individual career strategy: does society at large need to invent new mechanisms of retribution for creative work, given that the market mechanisms
that once supported musicians (unit sales of recordings) have been structurally undermined first by streaming and now further by AI?
As Pachet \cite{Pachet2025} states: ``The answer lies in finding new ways to recognize and value the unique contribution of each creator,
even as their work becomes part of a much larger, ever-evolving network.
This requires a shift in mindset--from seeing art as a fixed object to seeing it as a dynamic process, and from protecting ownership to fostering participation.''
We will see below some public possible answer to completely decouple some recognition and retribution policy from the traditional rights-based revenue framework.

\subsubsection{The Irish Experiment on Basic Income for the Arts}

One concrete tentative answer comes from Ireland. Since 2022, the Irish government has run a Basic Income for the Arts (BIA) pilot project \cite{BIA2026},
providing roughly 2,000 randomly selected professional artists and creative workers -- across all art forms, including music -- with an unconditional payment of \euro{325} per week,
regardless of their other income, for three years (later extended to early 2026).
A control group of applicants who were not selected received a token payment and participated in the same data collection, allowing for a genuine comparative evaluation.

The results, published by Ireland's Department of Culture in a 2025 cost--benefit analysis \cite{BIA2025}, are notable.
For every \euro{1} of public money invested in the scheme, Irish society received an estimated \euro{1.39} in return, once the value of increased creative output, audience engagement, psychological wellbeing, tax revenue, and reduced social welfare costs is accounted for
-- a net positive return rather than a pure transfer cost.
The scheme's gross cost over the pilot period was around \euro{105 million}, and the total estimated social and economic benefit exceeded \euro{100 million} on its own, even before factoring in the avoided welfare costs that brought the net fiscal cost down to roughly \euro{72 million}.
In other words, the resources and value generated by the supported artists' creative activity -- and the broader social benefits of that activity, including public engagement with their work
-- substantially exceeded the cost of the income guarantee itself. Following this evidence and a public consultation in which 97\% of respondents supported making the scheme permanent,
the Irish government has now folded a version of the BIA into permanent arts funding policy, though the precise shape of its long-term successor remains under discussion as of 2026.

The Irish experiment is, of course, a single national pilot and not a ready-made template -- it covers all art forms, not music specifically,
and questions remain about its scale, its selection mechanism (currently a random lottery among eligible applicants rather than a needs- or merit-based allocation), and its long-term fiscal sustainability if extended to a much larger population of artists.
But it offers a rare piece of empirical evidence, rather than mere advocacy, that decoupling at least part of musicians' and artists' income from the direct, atomized sale of their output -- exactly the model that streaming and now generative AI have eroded
-- can be not merely affordable but economically constructive for society as a whole. As AI continues to compress the market value of recorded music and most other forms of mass-produced creative content, schemes of this kind,
alongside the individual diversification strategies discussed above, are likely to remain part of an active and necessary policy conversation -- one that music education, by training the very people whose livelihoods are at stake, has a direct interest in following and informing.

\subsection{Other Issues}

We will now introduce some further issues and touch upon them rapidly, without extensive elaboration.
That would and may be the subject of other upcoming article(s).

\subsubsection{Music Quality Decrease}

On the apparent decline in musical quality: several converging and likely cumulative causes can be identified, independent of any single explanation.
These include the rise of functional music (optimized for mood and background use rather than attentive listening); a sheer increase in the volume of music produced,
diluting both listening attention and the economic incentive for refinement; the disappearance of the selective, curatorial role once played by record companies and producers before the music reached the public;
a shift of creative focus toward sound production\footnote{On that issue,
	see, e.g., \cite{PachetGitzinger2024}.}
and away from melodic and harmonic composition; and a broader uniformization driven by widely shared production tools, including DAWs and now generative AI, which tend to push output toward the statistical center of a learned style.

\subsubsection{Risk of Homogenization}

A recurring concern in both the musicological and the AI research literature is the risk of homogenization -- the convergence of musical production toward a smaller set of styles, textures, and forms.
The analysis of popular Western music over five decades by Serr\`a et al. \cite{Serra2012} documented a restriction in melodic variety and a homogenization of timbre well before the widespread adoption of AI.
The study on instrumental complexity by Percino et al. \cite{Percino2014} found that commercial success tends to follow, and in turn reinforce, stylistic simplification: as a style attracts more artists,
its instrumental variety grows; but as its commercial success becomes established, the pressure toward conformity increases.

Generative AI adds a new cause to this dynamic. As Doshi and Hauser \cite{DoshiHauser2024} found, AI-assisted creative outputs are more similar to each other than purely human-authored ones.
Shumailov et al.'s \cite{Shumailov2024} analysis of recursive training on AI-generated data -- \emph{model collapse} -- shows that models trained on their own outputs converge toward very low variance.
In music, a world in which streaming platforms train recommendation algorithms on AI-generated content, which in turn trains new generative AI models, which generate more content for recommendation, could drive rapid convergence toward a narrow sonic mean.

Music education has a crucial countervailing role to play here. By cultivating awareness of musical diversity -- historical, cultural, and stylistic --
and by teaching students to recognize and resist the seductive ease of stylistically conventional output (whether generated by AI or produced by human artists under commercial pressure), educators can help maintain the breadth and richness of musical life.

\subsubsection{Creativity: What Is It, and Can AI Have It?}

On AI and musical creativity: genuine creativity remains, in the author's view, presently unreachable by AI\footnote{See developed arguments in \cite{Briot2025}.}.
What current systems do is better described as exploration or discovery within a highly multidimensional, learned space of possibilities -- closer to search than to creation.
AlphaGo's celebrated move 37 against Lee Sedol is the emblematic case: an \emph{exploratory} move within an immense but bounded combinatorial space\footnote{It is important
	to note that AlphaGo AI foundational model is not exactly the same as current deep learning (artificial neural networks)-based generative AI systems.
	AlphaGo is based on another machine learning model, {\em reinforcement learning},
	where an agent learns to make optimal decisions through trial and error,
	by receiving rewards (or penalties) for every successive action, and therefore figuring out the best policy (strategy) over time.
	Current AI systems actually often combine both deep learning and reinforcement learning by using deep learning artificial neural networks models to represent the {\em policy}
	(also named the {\em actor}) or/and its evaluation (named the {\em critic}) of the learning agent, neural networks which are updated after each action and its reward.
	This is the case for AlphaGo (which also uses a Monte Carlo tree search -- MCTS) and,
	interestingly, also for current LLM-based chatbots such as ChatGPT and Claude,
	which, after the initial supervised training with the dataset of examples,
	apply {\em reinforcement learning from human feedback} (RLHF, in order to align the agent onto human ``values'')
	onto the LLM model/network considered as a policy, in general via {\em proximal policy optimization} (PPO) or a derivative.
	We stop here for these conceptual and technical important aspects, the important lesson to understand in the context of this paper is that current generative AI systems
	(which are based on artificial neural networks) fundamentally work through {\em extrapolation} from a (huge) set of examples, while reinforcement learning AI systems (such as AlphaGo),
	fundamentally work through {\em exploration}, therefore are more oriented towards {\em discovery},
	although the frontier tends to become thinner with advances on the artificial neural networks front, with, e.g., planning abilities through chain of thoughts and other meta-reasoning abilities.},
evaluated as optimal by the system's reward function, not by any aesthetic judgment the system itself possessed.
Several capacities still appear to be missing for a stronger claim of artificial creativity: a proper model of aesthetics (no clear, learnable function corresponds to ``good music'');
a robust and autonomous capacity for self-evaluation, including retrospective evaluation of one's own output; and a grounding of the generative process in lived, embodied human culture and experience.
In addition, current generative AI systems perform extrapolation from a corpus of pre-existing works, not a modeling of the creative process itself by which those works came to be.

This does not diminish the educational significance of AI creativity tools. Creativity in human education does not require that the tools used be themselves creative.
A piano or a paintbrush are not creative. What matters is that the tools are used in ways that stimulate genuine creative thought in the student.
AI composition assistants, used thoughtfully, can do this -- by generating unexpected harmonic possibilities for a student to evaluate and extend, by modeling a style for analysis and critique, or by challenging a student's assumptions about what ``sounds right.''

\section{Education Has to Change}
\label{sec:education-change}

\subsection{Online Platforms and the Transformation of Access: MOOCs and YouTube}

Before AI became the dominant topic in educational technology, a first wave of transformation was already underway through online platforms.
YouTube is perhaps the most consequential single development in informal music education of the past two decades.
Tutorials covering everything from basic guitar chords to advanced jazz harmony theory, from music production in Ableton to orchestration for film, are available for free in every major language.
This has had a genuine democratizing effect on access to music knowledge, particularly for learners in regions or socioeconomic situations where private instruction is inaccessible.

Massive Open Online Courses (MOOCs) have offered a more structured complement to informal YouTube learning.
Platforms like Coursera, edX, and specialized music platforms have made courses from institutions including Berklee College of Music and Juilliard available globally.
By 2024, the online music education market was estimated at several billion dollars and projected to reach nearly \$5 billion by 2030, growing on the back of improved Internet access, the widespread use of digital devices, and the demonstrated effectiveness of online learning in the COVID-19 context.

However, MOOCs face well-documented structural challenges: completion rates are notoriously low (typically under 10\%), sustained engagement is difficult to maintain, and the specifically musical dimensions of learning
-- embodied practice, real-time performance feedback, the social dynamics of ensemble playing -- are particularly resistant to the format of short video lectures and self-graded quizzes.
Research on social MOOCs, designed to mimic the learning community dynamics of workshops, small choirs, or hackathons, suggests that peer interaction and collaborative problem-solving are essential to effective music learning online, but these are difficult to engineer at scale.

For music education specifically, the online medium raises questions that go beyond completion rates. Learning an instrument, developing an ear, or internalizing a rhythmic feel requires sustained, embodied, and often supervised practice.
Platforms and pedagogical designs that take this seriously -- incorporating video analysis of student performance, automated feedback on pitch and rhythm, and structured peer review -- are more promising than simple lecture-video formats.

\subsection{AI-Based Training: Tools, Opportunities, and Limits}

The intersection of AI and music education is now a rapidly growing research field. A recent systematic literature review by S\'anchez-Jara et al.  \cite{SanchezJara2024} surveys the breadth of this activity, spanning domains from virtual and augmented reality to assistive technologies.
Four areas stand out as particularly consequential for how teaching and practice are likely to change in the coming years: personalized learning and intelligent tutoring, automated assessment and feedback, AI as a creative partner, and the evolving role of the teacher.

\begin{itemize}
\item \emph{Personalized learning and intelligent tutoring}. AI enables adaptive learning systems that adjust the difficulty, pacing, and content of instruction to the individual student's current level and learning trajectory.
In music, this means that a student practicing scales or sight-reading does not need to wait for a weekly lesson to discover whether they are ready to progress -- the system can analyze performance in real time and propose the next appropriate challenge.
Research by Ou et al.  \cite{Ou2025} found that AI-assisted practice applications demonstrably improved performance outcomes, self-efficacy, and self-regulated learning in music students, with tools analyzing pitch accuracy, rhythmic precision, and overall musical quality.
\end{itemize}

Specific tools are already in pedagogical use. SmartMusic and Tonara provide real-time feedback on pitch and rhythm during instrument practice, comparing student performance against a reference score.
Applications like Violin by Trala, with over 400,000 users across 193 countries, demonstrate that the appetite for AI-assisted instrument learning is large and global.
These tools are particularly valuable in contexts where access to expert instruction is limited -- they can provide the feedback loop that practice requires, even in the absence of a teacher.

\begin{itemize}
\item \emph{Automated assessment and feedback}. Automated assessment of musical performance has historically been limited to the most easily quantifiable parameters: pitch accuracy and rhythmic precision. Current research is pushing beyond these, toward the automated assessment of more nuanced expressive qualities -- phrasing, dynamics, articulation, stylistic idiosyncrasy. Deep-learning-based classification systems could provide some automated assessment
(see, e.g., \cite{Zhang2025} for some preliminary experiment), although the gap between such automated metrics and the rich evaluative judgment of an experienced teacher remains significant and is not trivially bridged.
\end{itemize}

For teachers, AI-powered learning analytics dashboards open the possibility of monitoring the practice and progress of many students simultaneously
-- tracking tempo stability, pitch accuracy, and harmonic comprehension across an entire cohort in ways that would be impractical with weekly one-to-one lessons alone.
This could shift the teacher's role from primary dispenser of feedback toward a more supervisory and interpretive function, reserved for the qualitative, contextual, and motivational aspects of instruction that machines cannot yet provide.

\begin{itemize}
\item \emph{AI as a creative partner in education}. The use of AI composition tools within music education contexts is perhaps the most provocative and contested area.
Generative AI can serve as a musical sparring partner: a student working on jazz harmony can ask a system to generate chord progressions in a given style for them to analyze or improvise over\footnote{See for instance
	potential offered by pioneering prototypes such as FlowComposer, which has been introduced in Section~\ref{sec:ai-in-picture}.};
a student studying counterpoint can ask for some harmonization of their melody by a ``virtual Bach\footnote{Such as
	DeepBach \cite{Hadjeres2017}.},''
which they then critique and compare to their own attempt. The Continuator itself, may be useful for training improvisation and offer a musician a way to reflect on their own play and patterns \cite{Pachet2006}.
\end{itemize}

Meanwhile, we should be also aware of limits of current AI technology. For instance, research by Doshi and Hauser \cite{DoshiHauser2024} on creative writing -- directly relevant by analogy --
found that generative AI assistance significantly raised the output quality of initially less creative individuals, though it did not significantly change the scores of initially highly creative ones.
Crucially, it also found that AI-assisted outputs were more similar to each other than purely human-authored ones, suggesting a homogenization risk.
In music education, this suggests that AI tools can be valuable scaffolds for students who are struggling to get started or who lack confidence, but that teachers must be attentive to the risk that AI-mediated creativity becomes a kind of mimicry rather than genuine exploration.

\begin{itemize}

\item The teacher's evolving role. All of these technological developments raise a fundamental question about the role of the music teacher.
None of the AI tools currently available -- and none on the near-term horizon -- can provide the full context-sensitive, emotionally attuned, and culturally grounded mentorship that expert human teaching provides.
Indeed, some very fundamental limitation of current AI technology is that it is not grounded in some subjective experience and esthetics.
In other words, although current AI technology can provide {\em scale} (feedback for many students simultaneously),
{\em consistency} (always-available practice partners),
and a certain kind of {\em patience}
(a system that does not tire of correcting the same rhythmic error for the five hundredth time!), it cannot provide yet the modeling of musical passion,
the transmission of a living performing tradition, or the sensitivity to a particular student's psychological state on a particular day.
\end{itemize}

As a summary, the most productive framing for AI in music education may be the same as in music creation: not autonomous replacement,
but \emph{intelligent assistance}. AI can relieve teachers of the most repetitive and quantifiable aspects of musical feedback, freeing them to concentrate on what only human teachers can do.

\subsection{Rethinking the Curriculum}

These changes collectively imply that music education curricula need to be reconsidered, not merely amended at the margins. Several directions seem to look clear:

\begin{itemize}
\item \emph{Production and DAW literacy} must become a standard component of music education at all levels, rather than a specialization.
The ability to work with loops, samples, audio effects, and a digital audio workstation is now a basic competency for any musician who intends to function in the contemporary music industry, regardless of genre.

\item \emph{AI literacy for musicians} must be cultivated -- not as a technical deep dive into machine learning architectures,
but as a practical and critical capacity: understanding what generative AI systems can and cannot do, how to use them as creative tools, and how to evaluate their outputs.
This includes the ability to distinguish between using AI as a {\em shortcut} (which raises legitimate questions of artistic authenticity) and using it as a {\em creative catalyst}.

\item \emph{Critical and contextual knowledge around rights, economics, and the social role of music} must become more prominent.
Students who do not understand the streaming economy, the legal status of AI-generated music, or the structural dynamics by which musical value is created and distributed are not equipped for the realities of a professional musical life.

\item \emph{Ensemble and performance} -- the embodied, social, and spontaneous dimensions of musical life -- must be defended and foregrounded precisely because they are the aspects most resistant to AI substitution.
The risk is that under economic pressure and the seductive convenience of digital tools, these dimensions are progressively marginalized in favor of what is more easily measured and delivered online.
\end{itemize}

\section{Conclusion}
\label{sec:conclusion}

The challenges outlined in this paper are real, deep, and interconnected. They cannot be addressed by simply adding an ``AI module'' to existing curricula, or by reactively banning AI tools from classrooms.
They require a more fundamental rethinking of what music education is for, what it should produce, and what kinds of musical intelligence it should cultivate.

Some directions seem clear. Music education must embrace the technological literacy that contemporary musical practice requires: DAW proficiency, AI tool fluency, understanding of the streaming economy and its legal landscape.
It must also resist the temptation to reduce musical education to what is most easily measured or most efficiently delivered online. The embodied, social, improvisatory, and culturally grounded dimensions of musical life
-- which are precisely what AI cannot provide and what streaming tends to efface -- must be defended with renewed clarity and urgency.

The relationship between human creativity and AI assistance offers a productive model. Just as the best AI-assisted composition tools do not replace the composer's intentionality but amplify it
-- as FlowComposer amplifies the musician's melodic and harmonic imagination, or the Continuator extends their stylistic voice into interactive dialogue
-- the best AI-augmented music education will not replace the teacher's wisdom or the student's growing musical self, but will extend and enrich both.

The final word belongs to the students. The musicians who will navigate this transformed landscape most successfully will be those who understand technology well enough to use it creatively\footnote{Young generations
	who are fluent with new technologies will be likely to adapt more easily to this new way of education.},
	who have a sufficiently rich musical culture to evaluate AI outputs critically, who have practiced their craft deeply enough to know what they want to say, and who have retained the curiosity and the capacity for surprise that any living creative practice requires.
	It is not certain that AI will make this easier. It is quite possible that it will make it more necessary.

\section*{Acknowledgments}

This text is the follow-up of my presentation to a round table on musical education in contemporary times
at the 7th Musical Education Meeting at Col\'egio Pedro II, in Rio de Janeiro, Brazil, on May, 29th, 2026.
I want to acknowledge the assistance of Claude (Sonnet 4.6, from Anthropic) to help me construct and refine this text, and FAPERJ (Brazil) for partial support of this study through a PV research fellowship.

\bibliographystyle{alpha}
\bibliography{Challenges_for_Musical_Education}

\end{document}